# 湍流手性：可逆系统的平衡与非平衡有螺系综

朱建州[①]

① 速诚基础与交叉科学研究中心, 南京，高淳，211316;
*联系人, E-mail: jz@sccfis.org

**摘要**　　研究基于修改 Navier-Stokes 方程而得到可逆守恒系统的有螺湍流研究思路和理论。讨论三维可压缩湍流统计力学分析及螺度控制力热声问题。构建了一个非平衡系综的有螺可逆系统。最后简要讨论工程应用。

**关键词**　　可压缩湍流，螺度，手性，统计拓扑流体力学，时间可逆系统，气动声学，螺控壁湍流力和热，量子声学

## 1　引言

湍流的手性文献可追溯到 Betchov 的螺度研究[1]。定义"螺度"为流场速度$u$与旋度$\omega = \nabla \times u$内积的空间积分, $\int h \, d^3x$, 所以我们这里把$h = u \cdot \omega$称为"螺旋度"。其它几何讨论，尤其是 Kelvin, Moreau 和 Moffatt 等人这条线的拓扑方面可参文献 [2]。

关于湍流的问题很多，本文关注的是，1) 什么是湍流动力学最基础和根本的？2) 应该如何去揭示湍流的基本性质？3) 这些基本的性质在复杂的自然现象和工程工况中所处的位置、有什么用、怎么用？我们都就聚焦的手性问题给出部分答案以抛砖引玉。

Navier-Stokes 湍流作为一个非线性数学/物理问题，人们首先会想到可积系统的"孤子"和"呼吸子"，以及混沌系统的"（奇怪）吸引子"。他们当然和研究湍流相关；但是当看到洪流滚滚风起云涌时，更让人思考大自由度系统的统计性质：均匀各向同性湍流这个"最简模型"首当其冲。

对于问题 1)，答案是非线性项相互作用的对称/守恒性质。三维湍流从大尺度到小尺度级串，越来越各向同性，最后为分子粘性所耗散。简单的思考可能会认为粘性在起主导作用；但是，要知道 Navier-Stokes 方程中已再无分子，如果我们把粘性项改为大尺度耗散作用项（数值模拟中称为"亚粘性/hypoviscosity"），难道三维湍流的级串图像就调转了吗？当然不是。下面我们介绍的李政道和 Kraichnan 工作给出的最基础和根本的非线性物理解释。

事实上，Burgers、Onsager、Hopf 和 Lee（李政

道)不约而同地都认准纯粹非线性项模态相互作用的动力学性质和统计结果下手。凭借能量守恒和相体积守恒（Liouville 定理），辅以遍历假设，获得平衡态统计解。Lee[3]对不可压缩流体和磁流体同时处理，明确提出各模态能量均分，为 Kolmogorov 的各向同性假设提供基础支持（这一结果不受后来的螺度修正影响。）Kraichnan[4]随后宣称弱可压缩流体平衡系综能量均分，并籍以讨论噪声与湍流的相互作用，与 Lee 和 Lighthill 比肩。Kraichnan 此后不论是专门计算纯非线性相互作用的统计平衡结果，还是做理论模型（如，1958 年的直接相互作用近似文章），都以非线性相互作用的守恒/对称性质贯穿始终：尤其于 1967 年发表二维流动绝对统计平衡结果并据此得到能量反向级串显著性推论；1973 年对李政道之前结果提出螺度修正。本文作者近年系统地对中性流体和等离子体模型进行这方面的计算分析（部分内容是对 Lee 政道、Kraichnan 和 Frisch 等人历史性结果的手性分解更新）。本文要介绍的是预测螺度压制压缩模的结果[5]。我们不一一罗列文献，读者可从给出的文献和专刊另文[6]获得更多信息。

上面介绍的工作是从动力学时间可逆的系统出发的。其相体积或 Liouville 测度的守恒保证了最终平稳分布的存在。由于理想 Euler 动力学是保相体积的，放弃它意味着可引入外源/汇，比如粘性耗散和外力驱动。而后者反而带来了附加其它守恒/对称性的自由度，可能通过适当调整源和汇之间的某种平衡以得到想要的守恒律，比如，拟涡能、能量、螺度，等等，甚至同时满足多个。而适当引入外源/汇以保证时间可逆性带来了严格精确分析的可能，比如存在且唯一的遍历测度、熵产生和李雅普诺夫指数间的关系等等[7]。

在分子动力学模拟中，为产生与正则或微正则系综在某种意义上等价的系综，实践中发明了各种具有可逆性质的技术，比如所谓 Gauss 方法[8]。这当然会让人想到在湍流问题中的尝试，比如，She（佘振苏）和 Jackson[9]，以及 Gallavotti 和 Cohen[8]就分别对波数矢量的每个波长壳上的能量和对总能量设计外力，使得他们不变，而且系统动力学是时间可逆的。Gallavotti 和 Cohen 提出涨落定理并发展时间可逆的湍流非平衡等价系综理论[8]。这些理论目前尚有局限。比如，都只能固定一个物理量。而二维湍流中动能和拟涡能，三维湍流中动能和螺度都具有对偶的重要性。

下文先借作者螺度抑制压缩模工作，表达统一经典和量子拓扑物态图景的尝试，然后介绍如何破解以往多约束超定的困难以引入能量-螺度双约束，建立新的时间可逆非平衡湍流统计系综。最后讨论应用。

## 2 湍流手性：可逆系统的平衡与非平衡统计力学视角

我们先引入（Fourier）螺旋分解：对于一个三维矢量场，比如速度 $\mathbf{u}$，可以通过进一步将横向场分解为（纯）左旋和（纯）右旋两部分来把 Helmholtz 分解更加精细化（[5]及其参考文献）：$\mathbf{u} = \mathbf{u}^\perp + \mathbf{u}^| = \mathbf{u}^+ + \mathbf{u}^- + \mathbf{u}^|$；考虑 Fourier 级数 $\mathbf{u(r)} = \sum_\mathbf{k} \hat{\mathbf{u}}_\mathbf{k} e^{\hat{i}\mathbf{k}\cdot\mathbf{r}}$（为方便，下面不言自明地依情况把 $\mathbf{k}$ 写在括号里作为变量的函数宗量或把它处理为变量的下指标的形式），我们有 $\hat{\mathbf{u}}(\mathbf{k}) = \hat{u}^+(\mathbf{k})\hat{\mathbf{h}}_+(\mathbf{k}) + \hat{u}^-(\mathbf{k})\hat{\mathbf{h}}_-(\mathbf{k}) + \hat{u}^|(\mathbf{k})\mathbf{k}/k$，其中 $\hat{i}\mathbf{k} \times \hat{\mathbf{h}}_{c_k} = c_k k \hat{\mathbf{h}}_{c_k}$：$\hat{i}^2 = -1$ 和 $c_\mathbf{k}^2 = 1$（$c_\mathbf{k} = "+"$ 或 $"-"$ 代表每个对应波矢 $\mathbf{k}$ 的模态绕 $\mathbf{k}$ 的两个相反的旋转方向：手性 --- 下面在不言自明的情况下，我们又是为了符号的方便会把这个 $\mathbf{k}$ 指标略去）。运用这种表示的好处在于三部分的螺旋度分别是正定、零和负定的。

描述不可压缩流动的 Navier-Stokes 系统为
$$\partial_t \hat{u}_\mathbf{k}^{c_\mathbf{k}} = \sum_{\mathbf{k}+\mathbf{p}+\mathbf{q}=0}\sum_{c_\mathbf{p},c_\mathbf{q}} \hat{C}_{\mathbf{kpq}}^{c_\mathbf{k} c_\mathbf{p} c_\mathbf{q}} \hat{u}_\mathbf{p}^{c_\mathbf{p}*} \hat{u}_\mathbf{q}^{c_\mathbf{q}*} - \nu k^2 \hat{u}_\mathbf{k}^{c_\mathbf{k}} + \hat{f}_\mathbf{k}^{c_\mathbf{k}}$$
(1)
其中 $\dfrac{\hat{C}_{\mathbf{kpq}}^{c_\mathbf{k} c_\mathbf{p} c_\mathbf{q}}}{c_q q - c_p p} = \dfrac{\hat{\mathbf{h}}_{c_p}^* \times \hat{\mathbf{h}}_{c_q}^* \cdot \hat{\mathbf{h}}_{c_k}^*}{2}$.

考虑粘性系数 $\nu = 0$ 和外力 $\mathbf{f} = 0$，由 $\hat{C}_{\mathbf{kpq}}^{c_\mathbf{k} c_\mathbf{p} c_\mathbf{q}}$ 对称性可以得到能量 $\mathcal{E} = \sum_{\mathbf{k},c_\mathbf{k}} |\hat{u}_\mathbf{k}^{c_\mathbf{k}}|^2$ 和螺度 $\mathcal{H} = \sum_{\mathbf{k},c_\mathbf{k}} c_\mathbf{k} k |\hat{u}_\mathbf{k}^{c_\mathbf{k}}|^2$ 在每个相互作用三波 $\{[\pm\mathbf{k},c_\mathbf{k}];[\pm\mathbf{p},c_\mathbf{p}];[\pm\mathbf{q},c_\mathbf{q}]: \mathbf{k}+\mathbf{p}+\mathbf{q}=0\}$ 中的细致守恒性质。

给定外力，在保持原方程非线性项的情况下，使系统变为时间可逆的方法就是让耗散项和外力的效果抵消。如果粘性力和外力处处完全互相湮灭，并且只修改粘性系数为依赖于波矢量的时变参数，那么 $\nu \to \mu_{\mathbf{k}}^{c_k} = (\hat{u}_{\mathbf{k}}^{c_k})^* \hat{f}_{\mathbf{k}}^{c_k} / (k^2 |\hat{u}_{\mathbf{k}}^{c_k}|^2)$（当分母为零时，如果这个波数上的外力不为零，则相应的动力学粘性系数 $\mu_{\mathbf{k}}^{c_k}$ 为无穷，否则也为零），则得到"自由的" Euler 方程。Fourier-Galerkin 截断就可以直观地理解为相应于每个截断的波矢量 $\mu_{\mathbf{k}}^{c_k}$ 为无穷。需要指出的是，这时对每个 $k$ 的左右旋两个模分别引入了动力学参数, 方程这时形式地对能量和螺度保持守恒。文献[9]让每个给定波长的波矢的壳上动能耗散和输入相抵消，即壳能量约束，来确定依赖于波长的时变参数；Gallavotti 及其合作者[8]进一步让总的动能（或其它可观测量，如拟涡能等）互相抵消，即更弱的全局/总能量约束，来确定全局的时变参数。事实上，容易验证，如果相抵消的是关于速度变量的二次非线性量，那么相应修改粘性系数为时变参数后的系统总是时间可逆的。对于相抵消的这个量来说，系统是守恒的。对于时间可逆的守恒系统，它可能展现出统计平衡态也可能展现出非平衡态，关键在于相流是否不可压（满足 Liouville 定理）。如果是，系统将趋于一个统计平衡态；否则，相空间收缩或膨胀，系统无法达到平衡，体现时间上的自发对称性破缺。

这个理论，尤其是关于只固定全局量的可逆系统的理论，有许多基本的问题需要回答：比如，在接受其基本思想的假设下，固定的物理量约束取什么？有什么区别？取一个还是多个？

## 2.1 可压缩湍流的各向同性极化问题绝对统计平衡分析一例

再引入 $\rho$ 和 $s$ 来表示流动的密度和声速，而热力学压强 $p$ 满足绝热关系 $p = s^2 \rho$。由于 $p$ 不依赖于温度（否则考虑等温情况），质量和动量方程从能量方程解耦。引入背景密度 $\rho_0$ 并做变换 $\rho = \rho_0 \exp\{\zeta\}$，和不可压缩情况一样，可压缩系统这时也满足 Liouville 定理。延续前述 Fourier 螺旋分解 $\hat{\mathbf{u}}^\perp(\mathbf{k}) = \hat{u}^+(\mathbf{k})\hat{\mathbf{h}}^+(\mathbf{k}) + \hat{u}^-(\mathbf{k})\hat{\mathbf{h}}^-(\mathbf{k})$，守恒的总能量在小涨落或低温下可近似为

$$\mathcal{E} = \sum_{\mathbf{k}} |\hat{u}^+|^2 + |\hat{u}^-|^2 + |\hat{u}^\parallel|^2 + s^2 |\hat{\zeta}|^2. \quad (2)$$

Kraichnan[4]由此得到能量均分原理，尤其是声模态和涡模态之间的均分。但是，正压（包括绝热）流动中螺度是守恒的【3】。因而，有必要引入 Lagrange 乘子 $\alpha$ 和 $\beta$ 而考虑运动常数 $C = \alpha\mathcal{E} + \beta\mathcal{H}$，以写出绝对统计平衡态的分布 $\sim \exp\{-C\}$。计算结果表明[5]，$\mathcal{H}$ 或 $\beta$ 不为零时不仅动能在左右旋间是极化的，体现出手性，而且涡模能量大于压缩模。

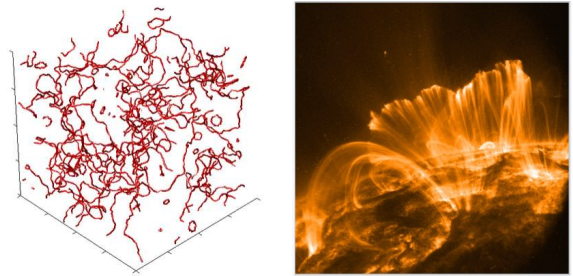

**图 1** 量子湍流中扭曲缠绕的涡线（左：https://en.wikipedia.org/wiki/Quantum_turbulence）和闪耀的日冕（右：https://en.wikipedia.org/wiki/Nanoflares#Nanoflares_and_coronal_heating）中互相缠绕拉伸磁场（被认为是冻结于等离子体物质）：具有类似可压缩湍流中的涡纽结重联释放自由能激发其它模态过程。

**Figure 1** Pictures (taken from wikipedia) of quantum vortex tanglement (left) and flaring coronal loops (right), indicating the similar scenario as in the compressible neutral fluid turbulence where the free energy is supposed to be released from the vortex knots and transferred to other modes by reconnection.

作者[5]根据上述讨论，预测可通过控制湍流中螺度来降噪。这里不重复深入细致的讨论，而是提供进一步的图像理解：相对于遍布的"小激波（shocklets）"[10,11]，我们也可以想象遍布的"小螺旋（helixlets）"和拥抱流体的涡绳或速度螺线。对于"紧固"固态物体，螺钉比无螺的钉子更好，非平凡的缠绕绳结比简单的好。气体和液体比固态物质更难以驾驭。而流动螺度度量流线螺紧性和涡绳纽结度，所以我们的结果说明，有螺度的这种"拓扑经典物态"压制了压缩模。同时，在这一思想指导下，由于（非线性）薛定谔方程可通过 Madelung 变换转化为流体方程，我们分析最近的一个非线性薛定谔方程（Gross-Pitaevskii 模型）数值模拟数据[12]，认为

他们的图六显示我们的结果同样适用于波色-爱因斯坦凝聚的波色气体的（更广意义下的）"拓扑量子物态"：当他们的螺度约束解除（t>4），压缩模、量子压力和声子模态迅速释放。这些结果对于等离子体物理也有直接或间接的启示（最后一节我们进一步评述与日冕加热的类比：图1。）当然，关于量子流体中与经典流体对应的螺度定义，学界还没有完全的共识，有待进一步研究。

## 2.2 手性非平衡"等效"统计系综的构造：同时控制动能和螺度的不可压缩时间可逆动力学

文献[5]通过时间可逆平衡系综计算给出前一节"螺控声"结果的同时，也给出了一个建立可压缩时间可逆非平衡系综的技术路径。这里我们对形式稍简单点的不可压缩情形展开讨论。

将 $c_\mathbf{k}$ 一致地固定为 $c$（'+'或'−'），两边同乘 $\hat{\mathbf{h}}_c e^{i\mathbf{k}\cdot\mathbf{r}}$ 并对 $\mathbf{k}$ 求和，然后将 $\nu$ 换为分别对应相应手征区的 $\mathcal{M}^c(t)$，方程（2）变为

$$\partial_t \mathbf{u}^c = \partial_t \sum_\mathbf{k} \hat{u}_\mathbf{k}^c \hat{\mathbf{h}}_c e^{i\mathbf{k}\cdot\mathbf{r}} = \mathcal{M}^c(t) \Delta \mathbf{u}^c + \\ + \sum_\mathbf{k} \hat{\mathbf{h}}_c e^{i\mathbf{k}\cdot\mathbf{r}} \sum_{\mathbf{k}+\mathbf{p}+\mathbf{q}=0} \sum_{c_p,c_q} \hat{C}^{cc_pc_q}_{\mathbf{kpq}} \hat{u}^{c_p}_\mathbf{p} \hat{u}^{c_q}_\mathbf{q} + \mathbf{f}^c. \quad (3)$$

进而令 $d\mathcal{E}/dt = 0$ 和 $d\mathcal{H}/dt = 0$，我们有

$$\mathcal{M}^\pm(t) = \frac{k^2|\hat{u}^c_k|^2 \sum_k k(\hat{f}^+_k \hat{u}^{+*}_k - \hat{f}_k \hat{u}^{c*}_k) \pm (\hat{f}^+_k \hat{u}^{+*}_k - \hat{f}_k \hat{u}^{-*}_k) \sum_k k^3|\hat{u}^\mp_k|^2}{\sum_k k^2|\hat{u}^c_k|^2 \sum_k k^3|\hat{u}^+_k|^2 + \sum_k k^2|\hat{u}^-_k|^2 \sum_k k^3|\hat{u}^-_k|^2}. \quad (4)$$

<u>（3,4）就是我们的三维不可压缩湍流能量-螺度约束的时间可逆动力学系综。</u>即便取相同的约束，构造时间可逆的动力学的方式不是唯一的。可以将原耗散项替换为任意关于 **u** 的其它奇次非线性函数，再进一步将其系数像上面一样替换为动力学（涡粘性）参数，也同样得到时间可逆的系统。取 Fourier 空间为其相空间，相体积是非守恒的，收缩率（动力学方程的散度：它不一定等于熵产生率）即为耗散率[8]。

我们的新系统既保持了原来 Gallavotti-Cohen 一套好的理论框架，又颠覆了原来约束任意而又单一的问题，与原始动力学建立更加密切的联系：要么会比原来他们的理论好，要么将其推到极端，接受更严格的检验，启发新思考。

上述方程（3,4）中引入两个动力学粘性参数构建三维可逆非平衡系综的思想可以很容易推广到二维情形，从而同时固定动能和拟涡能。

## 3 讨论

前面一节中我们通过湍流中体现手性的螺度的效应尝试回答了问题2)。这只是启示性的，尤其对于时间可逆的非平衡系综理论，还有大量的工作有待推进。下面讨论问题3)。

利用各种对称性/守恒律是数理科学研究重要手段，时间可逆性也使得非平衡系综更易于处理。尤其是很多非平衡统计的概念与平衡统计由此可以类比统一，理论上非常有潜力。

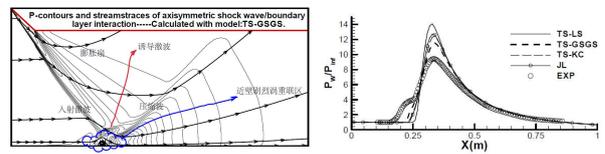

**图 2** 斜激波与柱体湍流边界层干扰数值模拟结果：左图通过云朵形状标出了用 TS-GSGS 模型模拟得发生剧烈涡重联的近壁区域，也就是对应于右图显示各种模型模拟和实验测量结果显示得壁面压力（系数）陡升（热流和摩阻系数图形类似）的壁面附近区域。

**Figure 2** Oblique shock interacting with the cylindrical turbulent boundary layer: In the left panel, the region of violent vortex reconnections simulated with a model called TS-GSGS is designated with a cloud, corresponding to the nearby region around the wall where the pressure coefficient goes up abruptly, similarly the heat flux and friction drag coefficients, in the right panel with results from simulations with various models and experiment.

近年来拓扑绝缘和超导量子物态备受关注，我们分析的更广意义下的拓扑经典物态同样有不平凡的性质。螺度的纽结论拓扑学解释与流体力学结合对经典和量子流动的螺控声效果给出了与紧固固体材料统一的图像，这不仅给人以统一性的美感，而且能带来巨大的工程应用价值：事实上，力热声是耦合的，我们的理论和进一步的数值实验结果（待发表）为螺度控制流动新思路提供了具体明确的指导。以壁湍流中激波边界层干扰[13]分离再附流动情况为例，可以认为这与量子湍流和解释日冕加热之谜的磁重连机制类似（图1）：涡结构在这里发生剧烈的重联过程，将原来涡纽结中的自由能释放，转化为热能和压力能等，从而可见壁面压力（图2）和热流陡升。可以从

我们的结果推论，如果控制螺度使得涡结构更稳健，重联减少，那么力和热的性能将可以大大改善。这在一定程度上也可通过涡绳纽结的螺度界定能量弛豫机制[14]来理解：给定能量，螺度越大，弛豫的余地就越小，反而言之涡绳将流体束缚得越紧。

对于螺控气动声，因为简单的操作会带来额外难以控制的噪声源，为了达到净降噪效果，还需要追溯到理论细节和大量投入的应用研究。2.1 节讨论的 Bose-Einstein 凝聚类似的螺度效应[12]，控制其中的螺度在高新技术领域的应用也有可观的前景。

**致谢**　向评审人和对该文有帮助的人士表示谢意.

# Equilibrium and non-equilibrium time-reversible dynamical ensembles relevant to chiral turbulence

Zhu J Z[1]

[1] *Su-Cheng Centre for Fundamental and Interdisciplinary Sciences, Gaochun, Nanjing 211316, China*

Ideas and theories of turbulence based on modifying the Navier-Stokes equation, to obtain equilibrium and non-equilibrium time-reversible dynamical ensembles relevant to helical turbulence, are presented. Discussions of controlling helicity to control the aerodynamic force, heat and noise are presented, together with the compressible turbulence relevant statistical mechanics analysis. A helical time-reversible system for nonequilibrium dynamical ensemble is constructed. Applications are also remarked.

**Compressible turbulence, helicity, chirality, statistical topological fluid mechanics, time-reversible system, aeroacoustics, helically-controlled force and heat in wall-bounded turbulence, quantum acoustics**